\documentclass[aps,pra,superscriptaddress,showpacs,twocolumn]{revtex4}
\usepackage{graphicx,bm,amsmath,amsfonts,amssymb,color,subfigure}

\newcommand{\be}{\begin{equation}}
\newcommand{\ee}{\end{equation}}
\newcommand{\bea}{\begin{eqnarray}}
\newcommand{\eea}{\end{eqnarray}}

\begin{document}

\title{Unconventional modes in lasers with spatially varying gain and loss}

\author{Li Ge}
\affiliation{Department of Electrical Engineering, Princeton University, Princeton, New Jersey 08544}
\author{Y.~D.~Chong}
\affiliation{Department of Applied Physics, Yale University, New
  Haven, Connecticut 06520}
\author{S.~Rotter}
\affiliation{Institute for Theoretical Physics, Vienna University of Technology, A--1040 Vienna, Austria, EU}
\author{H.~E.~T\"ureci}
\affiliation{Department of Electrical Engineering, Princeton University, Princeton, New Jersey 08544}
\author{A.~D.~Stone}
\affiliation{Department of Applied Physics, Yale University, New
  Haven, Connecticut 06520}

\begin{abstract}
We discuss a class of lasing modes created by a spatially inhomogeneous gain profile. These lasing modes are ``extra
modes", in addition to, and very different from, conventional lasing modes, which arise from the passive cavity resonances.
These new modes do not have high intensity across the entire gain region, but instead are localized at the gain boundary and throughout the gain-free region. They are surface modes, originating from the transmission resonances of the gain-free region. Using an S-matrix description we connect these surface modes to the lasing modes in $\cal PT$-symmetric (balanced gain-loss) cavities.
\end{abstract}
\pacs{42.25.Bs, 42.25.Hz, 42.55.Ah}

\maketitle

Generally in the theory of lasers, the observed lasing modes arise from the modes of the passive cavity, which are determined by the geometry and dielectric function of the cavity in the absence of gain. The cavity mode that evolves into the first lasing mode is selected because it has a relatively long lifetime and a large spatial and spectral overlap with the gain medium. Effects such as ``line-pulling'' \cite{Lamb_book,Haken_book,tureci_pra06}, self saturation, and mode competition can cause lasing frequencies to shift away from their passive cavity values, but typically each of the first few lasing modes can be identified as evolving directly from a single passive cavity mode.

There are laser systems in which the spatial distribution of the gain medium also plays a significant role
in determining properties of the lasing mode; two examples are gain-guided stripe lasers \cite{stripe} and gain-coupled
distributed feedback (DFB) lasers \cite{dfb1,dfb2}. In gain-guided semiconductor stripe lasers, lateral confinement is created by the injected carriers, which produce a weak graded index profile perpendicular to the direction of propagation of the light.  The confinement eliminates transverse modes and reduces the threshold current for lasing.  In gain-coupled DFB lasers an imaginary index grating is imposed at the Bragg frequency, which, though small, strongly affects the relative lasing thresholds of the passive cavity due to its periodic nature.  In both cases the resulting lasing modes emit at approximately the resonance frequencies of the passive cavity and are still in one-to-one correspondence with them.

Recent numerical studies in one-dimensional (1D) random lasers \cite{Andreasen_OptLett,Andreasen_pra} demonstrated the existence of lasing modes created by spatially varying gain which do not occur near passive cavity resonance frequencies and do not evolve smoothly out of passive cavity modes. Instead they appear at bifurcations in the solution set of the threshold lasing equations as a parameter, such as the length of the gain-pumped region, is varied.  They are ``extra modes", not in one-to-one correspondence with resonances of the passive cavity. These solutions were observed under conditions of weak scattering and a spatially inhomogeneous gain profile, with the signature of a stronger spatial localization when compared to the passive modes of the random cavity. It was not clear what role these three physical conditions --- weak scattering, randomness, and spatially inhomogeneous gain --- played in giving rise to these modes. In the present paper, we show that the phenomenon is primarily due to the spatially inhomogeneous gain profile and does not require randomness.

Using a 1D cavity partially filled with a pumped gain medium as an example, we show that these nonconventional modes are closely related to transmission resonances that occur as light propagates from the gain region through the gain-free region to
the external freely-propagating region.  They are ``surface modes", strongly-peaked at the boundary between gain and gain-free regions, and are not modes associated with an effective cavity created within the gain region.

A related problem of unconventional lasing modes has been studied recently \cite{Longhi,CPALaser}, where the cavity is divided not between gain and gain-free, but between gain and loss such that parity-time ($\cal PT$) symmetry is preserved; the combination of parity and time-reversal (which interchanges gain and loss) is a symmetry of the problem.  Interestingly such cavities, with no net single-pass gain, can nonetheless lase. Such cavities at threshold are called CPA-lasers because they can simultaneously function as laser amplifiers and as coherent perfect absorbers (CPAs) \cite{cpa_prl,cpa_science} at the lasing frequency \cite{Longhi,CPALaser}.
These lasing modes also bear no simple resemblance to passive cavity modes and are not in one-to-one correspondence to them. In the last section of this paper we show that these CPA-laser states are closely related to the surface modes, both of which are accompanied by an avoided crossing of the poles of the corresponding S-matrix.

\section{Spatially inhomogeneous gain}
\label{sec:partialgain}

For simplicity, we will focus on the unconventional lasing modes found in partially pumped one-dimensional cavities in which the electric field may be treated as a scalar. Let $\epsilon_c(x)$ denote the (real) dielectric function of the passive cavity. We will only deal with threshold lasing modes here and follow the approach of the Steady-state Ab Initio Laser Theory (SALT) presented in references \cite{tureci_pra06,SPASALT}.  At threshold, modal interactions can be neglected, and each lasing mode $\Psi_\mu(x)$ and its frequency $\omega_\mu$ can be derived from the following equation \cite{SPASALT}
\be
\left[\nabla^2 +  (\epsilon_c + \eta_\mu F(x))k_\mu^2\right] \, \Psi_\mu(x,\omega_\mu) = 0,
\label{eq:TLM}
\ee
with the boundary condition that the solutions be purely outgoing at $x \to \pm \infty$. Here $k_\mu\equiv\omega_\mu/c$ and $F(x)$ describes the spatial gain profile which is zero in gain-free regions \cite{bibnote0}. For a fixed frequency $\omega$, this equation has solutions at discrete complex values of $\eta_\mu(\omega)$; by varying $\omega$ one finds pairs of $(k_\mu,\eta_\mu)$ which represent threshold lasing frequencies and electric susceptibility of the gain medium. The corresponding $\Psi_\mu(x,\omega_\mu)$, determined by (\ref{eq:TLM}), are the threshold lasing modes on which we will focus in our discussions below. Note that all $\eta_\mu$ must have a negative (amplifying) imaginary part.

For simplicity, we will adopt the ``linear gain model", in which we assume that the pumped gain medium only alters the imaginary part of the index in the cavity, i.e. $(\epsilon_c + \eta_\mu F(x))^{-1/2} \approx n(x)\equiv n'(x)  + in^{\prime\prime}(x) $, where $n'(x)=\sqrt{\epsilon_c(x)}$, $n^{\prime} (x) = -n_i F(x)$, and $n_i$ is a small real positive constant.  In this approach the parameters to be varied to reach threshold are the lasing frequency and $n_i$. The thresholds are ordered, beginning with the lowest value of $n_i$, corresponding to the first lasing mode to turn on as the pump is increased. The infinite set of threshold solutions are specified by $(\omega_\mu,n^{(\mu)}_i)$, but in a real laser once the first mode turns on it affects the thresholds and frequencies of other modes via non-linear effects.  These can be taken into account by the full SALT \cite{SPASALT}, but are neglected here.

\begin{figure}
\begin{center}
\includegraphics[clip,width=0.8\linewidth]{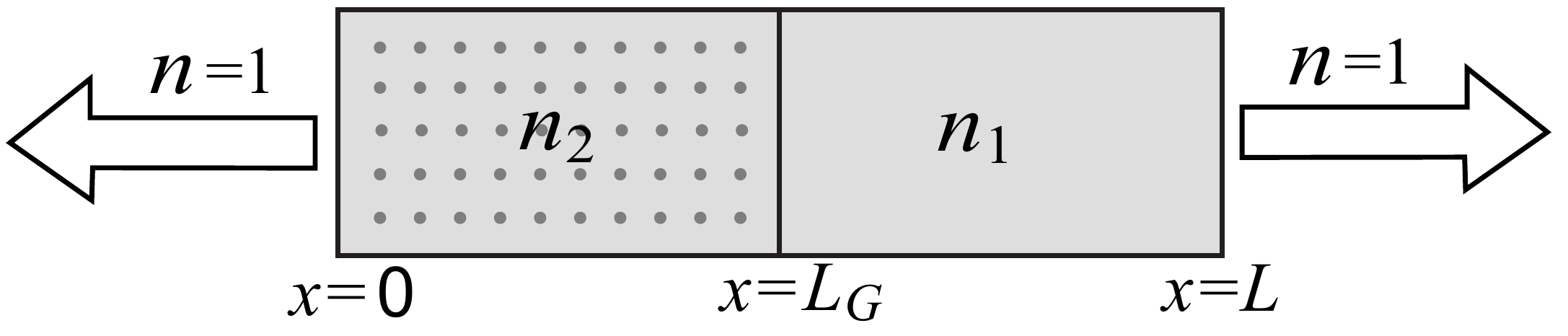}
\caption{Schematic diagram of a two-sided 1\textsc{D} edge-emitting laser.  The cavity occupies the region $0 \le x \le L$.  Gain is present in the dotted region ($0 \le x \le L_G$).}
\label{fig:cavity}
\end{center}
\end{figure}

In the ``partial gain'' system shown in Fig.~\ref{fig:cavity}, the gain medium is uniformly distributed in the left part ($0\le x \le L_G$) of a cavity of length $L$ and dielectric constant $\epsilon_c(x)=n_1^2 >1$. Emission occurs through two side walls at $x=0$ and $x=L$. Outside the cavity $\epsilon$ is taken to be unity. As noted, we assume that the refractive index of the gain region, $n_2$, is of the form $n_2 = n_1 - in_i\, (n_i>0)$.

\begin{figure}
\begin{center}
\includegraphics[width=0.8\linewidth]{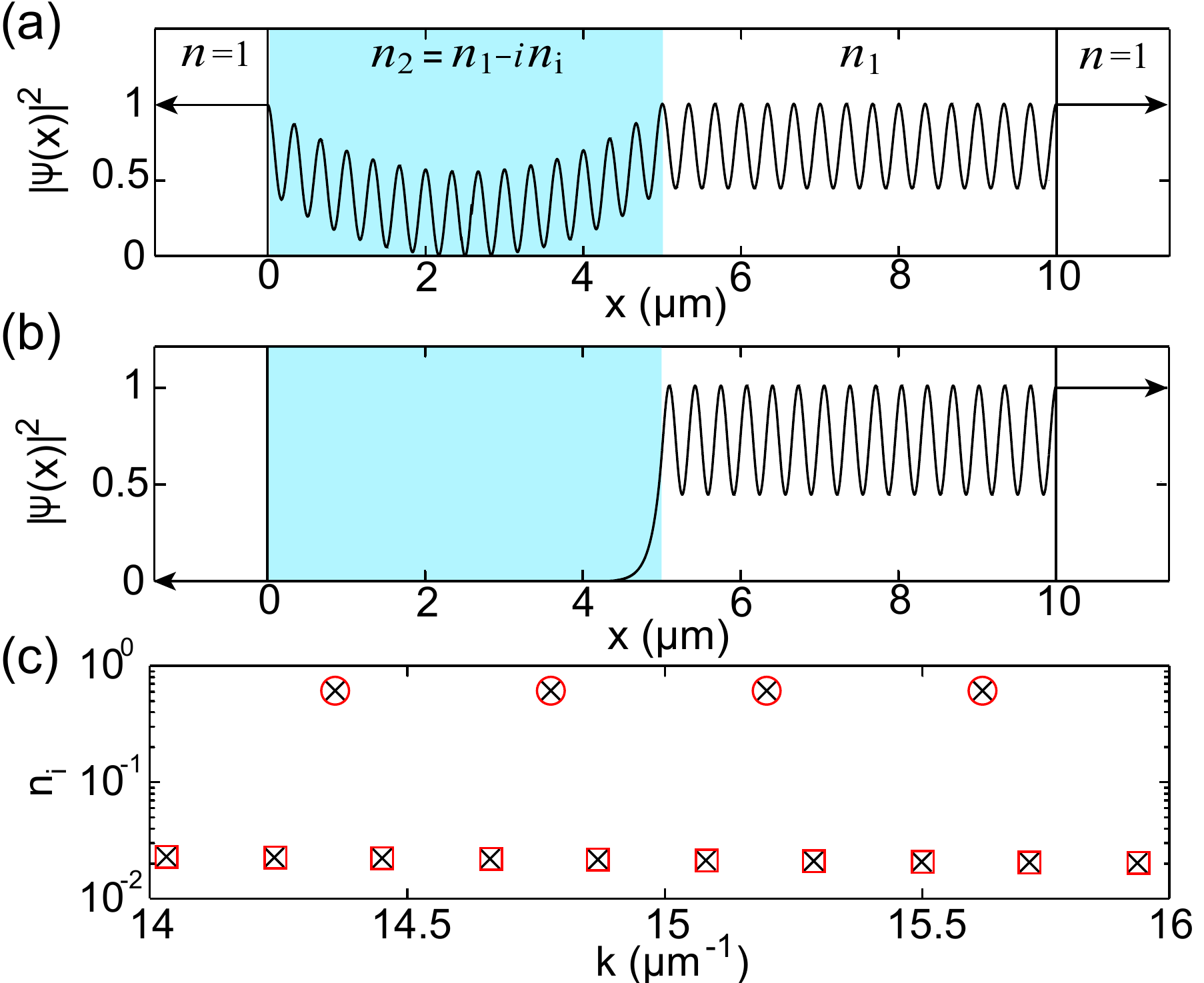}
\caption{(Color online) (a) Mode profile of a conventional mode ($k=6.2886\, \mu\textrm{m}^{-1}$, $n_i=0.0511$) in a one-dimensional cavity of length $L=10\, \mu\textrm{m}$ and index $n=1.5$. It is amplified symmetrically in the gain medium, which resides in the left half of the cavity (light blue region). (b) Mode profile of a surface mode ($k=6.4013\, \mu\textrm{m}^{-1}$, $n_i=0.6124$) localized inside the right edge of the gain regime. (c) The conventional modes (squares) and surface modes (circles) with $k \in [14, 16]\,\mu\textrm{m}^{-1}$. The crosses are the approximations given by Eq.~(\ref{eq:k1}-\ref{eq:k2}). The spacing of the conventional modes is defined by the whole cavity while the larger spacing of the surface modes is defined by the gain-free region. }
\label{fig:2sol}
\end{center}
\end{figure}

The solution to (\ref{eq:TLM}), with outgoing boundary conditions at $x=0$ and $L$, satisfies the following equation:
\bea
\frac{\tan[\theta_2+n_2kL_G]}{n_2} + \frac{\tan[\theta_1+n_1k(L-L_G)]}{n_1}=0,&& \label{eq:M22=0} \\
\theta_p \equiv \frac{\pi}{2} + \frac{i}{2}\ln\left(\frac{n_p+1}{n_p-1}\right),\;\; p = 1,2.&&
\eea
For large $kL_G$, this equation has two distinct sets of solutions. The first set can be found approximately by setting
 $n_2 = n_1$ everywhere in Eq.~(\ref{eq:M22=0}), except in the term $n_2k\,L_G$. This yields the familiar equally-spaced lasing frequencies, equal to the real part of the resonance frequencies of the passive cavity:
 \be
 \omega = c\frac{m\pi}{n_1L}, \quad m=0,1,2,... \label{eq:k1}
 \ee
As can be seen in Fig.~\ref{fig:2sol}(a), these modes have an oscillatory form throughout the entire cavity, but overlaid upon this oscillation is a clear amplification trend towards both ends of the gain region. The corresponding solutions for $n_i$ are
\be
n_i = \frac{n_1}{m\pi}\frac{L}{L_G}\ln\left(\frac{n_1+1}{n_1-1}\right)>0.\label{eq:n_i1}
\ee
This is precisely the value obtained by balancing the amplification of the mode in the gain region against the outcoupling loss due to Fresnel scattering at the two ends of the cavity.  The threshold amplification decreases with increasing $n_1$ and inversely with $k L_G$, tending to the standard limit for uniform pumping as $L_G \to L$.

In addition to these conventional modes, which are simply related to the passive cavity resonances, the partial gain system supports a second set of solutions, which we call ``surface modes''.  An example of a surface mode is shown in Fig.~\ref{fig:2sol}(b); the mode profile is dominated by anisotropic exponential growth within the gain region towards the boundary with the gain-free region.  The surface modes have a much higher threshold $n_i$ than the conventional modes; in the large $n_ikL_G$ limit their thresholds are given by
\be
 n_i \approx (n_1-1)\sqrt{n_1} >0. \label{eq:n_i2}
\ee
independent of $L_G$ and $k$.
The corresponding lasing frequencies are also equally spaced:
\be
 \omega \approx c\frac{m\pi + \arctan(\sqrt{n_1})}{n_1\,(L-L_G)},\quad m=0,1,2,... \label{eq:k2}
\ee
but the spacing is inversely proportional to $(L-L_G)$ rather than $L$. This dependence is very significant.  First, one sees that these modes are pushed off to infinite frequency as $L_G \to L$, so they do not exist in the uniformly pumped system. Second, if these modes were associated with an effective cavity created by
the gain region, one would expect a different dependence, i.e. $\omega \sim {L_G}^{-1}$.  Thus, surface states must arise from a different physical mechanism, which we will elucidate in the next section.

Fig.~\ref{fig:2sol}(c) shows $\{k,n_i\}$ for the two sets of modes, in a cavity of length $L = 10\, \mu\textrm{m}$ and $n_1=1.5$; notice the uniform but different spacing of the two sets.  For these parameters, the thresholds of the surface modes are almost two orders of magnitudes larger than the conventional modes unless $kL_G$ approaches unity. In general we can show that clearly distinct surface and conventional modes exist until a crossover wave number, $k_c$, given by
\be
k_c L_G =
\frac{\sqrt{n_1}}{(n_1-1)}\ln\left(\frac{n_1+1}{n_1-1}\right), \label{eq:k_c}
\ee
i.e. all the way down to near the lowest passive cavity frequency unless $n_1 - 1 \ll 1$ or $L_G \ll L$.
This explains why surface modes have not been observed experimentally to date and were first seen in
simulations of random lasers with very weak index variation around unity.

\begin{figure}
\begin{center}
\includegraphics[width=0.8\linewidth]{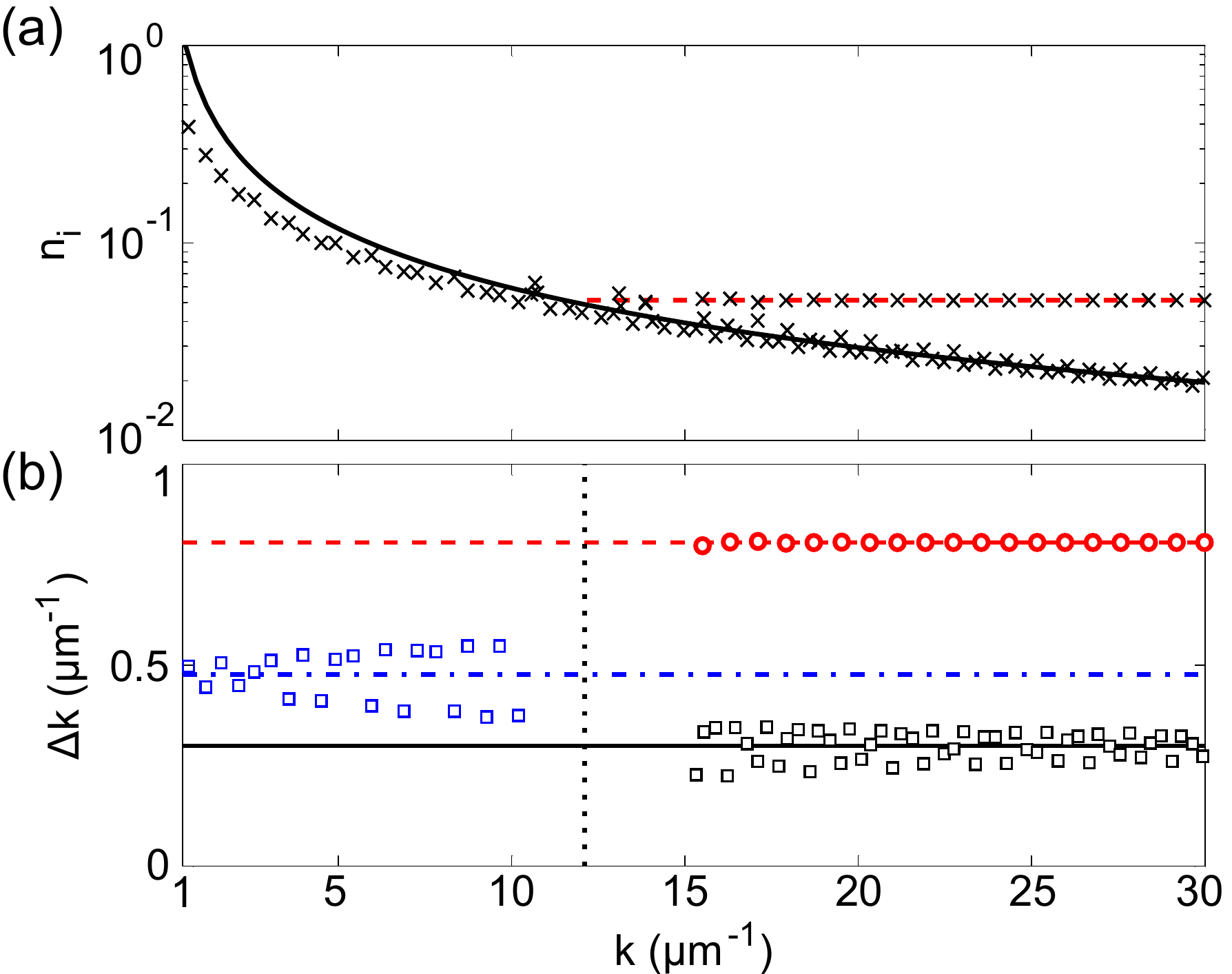}
\caption{(Color online) (a) Conventional modes and surface modes (crosses) in a one-dimensional cavity of $L=10\, \mu\textrm{m}$, $L_G=2\pi\,\mu{m}$ and index $n=1.05$. Solid and dashed curves indicate the approximations (\ref{eq:n_i1}) and (\ref{eq:n_i2}), respectively. (b) Spacing of the conventional modes (squares) and surface modes (circles). Solid, dashed, dash-dotted lines indicate the spacing determined by $L$, $L-L_G$ and $L_G$, respectively. The crossover of the squares near $k_c=12.11\, \mu{m}^{-1}$ (Eq.~(\ref{eq:k_c}); dotted line) marks the transition of the conventional modes from those of the passive cavity to those of the effective ``gain cavity''. Only data sufficiently far away from $k_c$ are shown where both the conventional modes and the surface modes show regular frequency spacing. }
\label{fig:n1_05}
\end{center}
\end{figure}

When $n_1-1 \ll 1$ we can study modes both above and below $k_c$ (Fig. \ref{fig:n1_05}). Above $k_c$ there are two sets of distinct modes, with the properties just discussed. As $k$ approaches $k_c$ from the higher frequency side, the fluctuation of the frequency spacing of the conventional modes ($\Delta k^{(c)}$) increases gradually (see Fig.~\ref{fig:n1_05}). As $k$ becomes smaller than $k_c$, the surface modes and some of the conventional modes do not exist any more, and $\Delta k^{(c)}$ has an even bigger fluctuation. However we find that the average of $\Delta k^{(c)}$ is no longer $\pi/n_1L$ but rather $\pi/n_1L_G$. Thus below $k_c$ the lasing modes do behave approximately as modes confined by the ``gain cavity".  This is perhaps not too surprising because the threshold normally increases as $k$ becomes smaller and the index discontinuity due to the imaginary index boundary can become more important than that due to the real index jump, especially in the limit $n_1 -1 \ll 1$. Some lasing modes in sub-wavelength plasmonic waveguides may be attributed to these effective ``gain cavity'' modes \cite{Hill}.

\section{Resonant tunneling interpretation}
\label{sec:resonantTunneling}

The approximation (\ref{eq:n_i2}) for the threshold of the surface modes was obtained for $n_i kL_G \gg 1$, i.e.~for amplification length much smaller than $L_G$.  In this limit, we can ignore the left edge of the gain region in Fig.~\ref{fig:cavity}.  The partial gain system now simplifies to a slab of index $n_1$ and length $(L-L_G)$ connected to two semi-infinite media, with indices $n_2 \in \mathbb{C}$ on the left and $n=1$ on the right. The scattering resonances for this system occur
at the frequencies given by (\ref{eq:k2}), and require that the refractive indices of the semi-infinite gain region and
the gain-free region satisfy
\be
\left(\mbox{Im}[n_2]\right)^2 = (\mbox{Re}[n_2]-1)(n_1^2-\mbox{Re}[n_2]). \label{eq:resonanceCondition}
\ee
For $n_2 = n_1 - in_i$, this is equivalent to (\ref{eq:n_i2}), defining the threshold value of $n_i$ for the surface modes.
We thus conclude that surface modes occur when light escapes through the gain-free region by resonant tunneling. It is therefore unsurprising that the spacing of the surface modes is inversely proportional to $(L-L_G)$ rather than $L_G$; their frequencies are related to the resonances of the gain-free region.  This also explains why the surface modes are never concentrated on the left boundary of the gain region.

For the system with finite gain and gain-free regions, one can understand the existence of
surface modes at the resonances of the gain free region by regarding the gain region as a uniformly pumped cavity enclosed by two very different mirrors.  The left ``mirror", with reflection coefficient $r_1$, is just the dielectric interface between vacuum and $n_2$, giving rise to simple Fresnel reflection, which is independent of frequency if $n_2$ is fixed. The right ``mirror", with reflection coefficient $r_2$, is the entire
extended gain-free region, with multiple reflections from both the $n_2 \to n_1$ interface, and the $n_1 \to 1$ interface.  Thus the reflectivity $r_2$ is strongly frequency-dependent with near-zeros at the frequencies of the resonances of
semi-infinite gain model just discussed.

The mode profile within the gain region can be decomposed into two amplified waves traveling in opposite directions:
  \be
  \Psi(x) = a_> e^{in_2kx} + a_< e^{-in_2k(x-L_G)}.
  \ee
At lasing threshold we have the usual round trip phase and amplitude condition for a 1D resonator:
$r_1r_2\exp(2i n_2k L_G) = 1$.  Using this relation and the definition of $r_1$, we find the ratio
of the right and left traveling waves to be:
 \be
 \left|\frac{a_>}{a_<}\right| = |r_1| \,e^{n_i k L_G} = \sqrt{\left|\frac{r_1}{r_2}\right|} \label{eq:ampRatio}.
 \ee
But for the surface modes $|r_2 (\omega)| \ll 1$, the lasing modes are dominated by right-moving waves, leading to the large asymmetry seen in Fig.~2. In contrast, $|r_2| \sim |r_1|$ for conventional modes, resulting in a roughly symmetric profile within the gain region.  From this point of view it is clear that in general the surface modes will have much higher thresholds since they occur when one of the ``mirror reflectivities" is strongly reduced compared to the conventional modes.

\begin{figure}
\begin{center}
\includegraphics[width=0.8\linewidth]{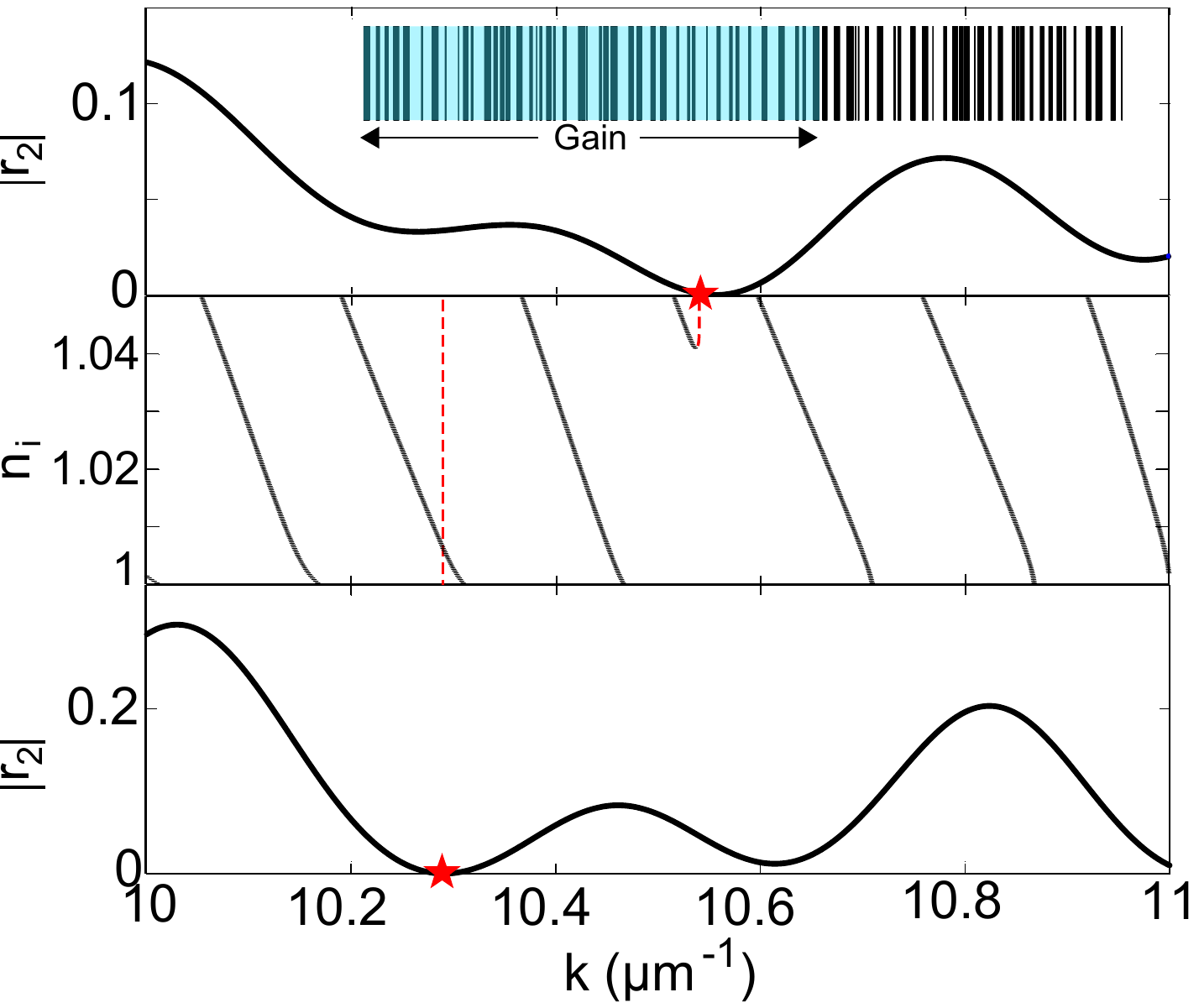}
\caption{(Color online) Center panel: Shift of threshold lasing frequencies when the refractive index of the dielectric layers {\it inside} the gain region is varied. The frequencies of the surface modes (vertical dashed lines) are stable despite the index change. Inset: The system under consideration is a weakly scattering 1D random system examined in Refs.~\cite{Andreasen_OptLett,Andreasen_pra}. It consists of 81 layers of dielectrics of averaged width $\langle d\rangle=0.15 \mu{m}$ separated by air gaps, with a total length of $24.1\mu{m}$. The gain (the light blue region) is uniformly distributed upto the the 47th layer from the left. Upper and lower panels: Reflection coefficient as a function of $k$ when the imaginary part of the index inside the gain region corresponds to the values of the two surface modes ($n_i = 0.0265$ and $0.4094$), respectively. The stars indicate the frequencies of the two surface modes $k = 10.29,\,10.54\,{\mu}m^{-1}$.}
\label{fig:kShift}
\end{center}
\end{figure}

Having clarified the physical picture of these unconventional modes for a simple partial gain system,
we now verify that it also applies to
the unconventional modes in weakly scattering 1D random media, discovered in Refs.~\cite{Andreasen_OptLett,Andreasen_pra}. We consider a random cavity with weak index variation as shown in Fig.~\ref{fig:kShift} (specific parameters are given in the figure caption). Nine lasing modes are found in the interval $k \in [10,11]\,\mu\textrm{m}^{-1}$ shown, two of which are found to be surface modes associated with resonant tunneling through the gain-free region.  This is determined by the following procedure: we smoothly reduce the scattering strength {\it inside} the gain region by reducing $n_1$, defined by
$\text{Re}[n(x)] = 1 + (n_1-1) f(x)$ in which $f(x)$ is a stepwise function with values $0$ and $1$. The frequencies of the conventional lasing modes blueshift systematically, while the frequency of the two surface modes barely change (vertical lines in the center panel). This finding confirms that the surface modes are determined by the properties of the gain-free region and have negligible dependence on the gain region. To further confirm this interpretation, we calculate the reflection coefficient of the random gain-free region, for light in the last gain layer connected to it. The choice of the gain coefficient, $n_i$ is different for each of the two modes, as they have different thresholds. This leads to the two plots shown in the upper and lower panels of Fig.~\ref{fig:kShift}. The lower panel is for the lower frequency surface mode, and as expected shows a near-zero of the reflection coefficient at the frequency of this mode; the upper panel is for the higher frequency surface mode and has the same behavior.

An additional effect seen explicitly in the center panel of Fig.~\ref{fig:kShift} is an inverse bifurcation of the solution set as a parameter is varied, in this case the overall amplitude of $n_1$ in the gain region. The surface mode, with $k=10.54\,{\mu}m^{-1}$, annihilates with its closest conventional mode at $n_1=1.041$, where their thresholds become identical.  This need not happen at all frequency crossings as shown by the other surface mode in the figure.  Similar bifurcations (in the forward direction) occur throughout the threshold lasing spectrum as $L_G$ is reduced from $L$, producing the surface modes.

\section{S-matrix Picture and Lasing in $\cal PT$-symmetric cavities}
\label{sec:Smatrix}

In this section, we study the partial gain system from the point of view of the scattering matrix (S-matrix). As we shall see, this reveals a unified description linking conventional and surface modes to laser-absorber (CPA-laser) modes of $\cal PT$-symmetric cavities \cite{CPALaser}.

The S-matrix of a 1D cavity is defined by
\begin{equation}
  S(n(x),\omega) \left(\begin{matrix} B \\ C \end{matrix}\right)
  = \left(\begin{matrix} A \\ D \end{matrix}\right),
\end{equation}
where $B,C(A,D)$ are the complex amplitudes of the incident (scattered) waves from the left and right side of the cavity. Although physical solutions are associated with real $\omega$, it is standard to analytically continue $S$ to complex values of $\omega$.
Scattering resonances correspond to purely outgoing boundary conditions on the S-matrix ($B = C = 0$) and occur when
the S-matrix has a pole in the complex plane. In a lossless cavity $(n(x) \in \mathbb{R})$ the poles of the S-matrix are required by current conservation and causality to occur only at complex values of $k$ in the lower half plane (with a negative imaginary part). For the simple dielectric cavity of Fig.~\ref{fig:cavity}, in the absence of gain, the resonance wave numbers are
\begin{equation}
  k^{(0)}_m = \frac{\pi m}{n_1 L} - \frac{i}{n_1 L} \ln\left(\frac{n_1 + 1}{n_1-1}\right),\,(m = 1, 2, \cdots);
\end{equation}
the imaginary part reflects the cavity outcoupling loss and defines the threshold values of $n_i$ when gain is added.

The real part of $k^{(0)}_m$ defines frequencies at which linear scattering is resonantly enhanced, but the outgoing solution
at the complex wave number $k^{(0)}_m$ is not itself a physically realizable state as it corresponds to a non-conserved photon flux {\it outside} the cavity (both in the upper and lower half planes).  Poles precisely on the real axis {\it do} correspond to physical states, and we give them the different name of ``threshold lasing modes" (TLMs) \cite{SPASALT} to distinguish them from other resonances.  It is the TLMs of the partial gain system which we studied in the earlier sections of the paper.

\begin{figure}
\begin{center}
\includegraphics[width=0.8\linewidth]{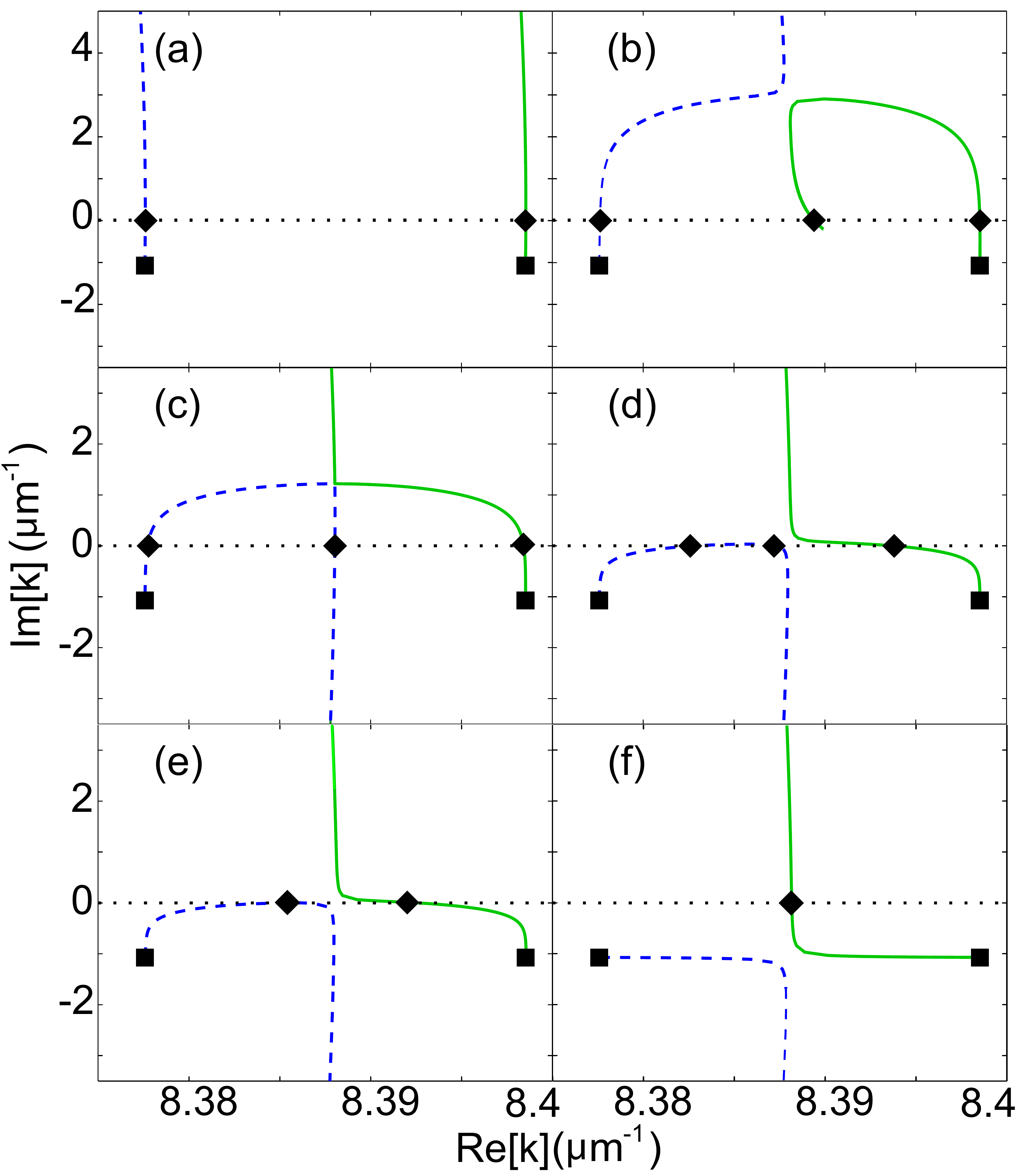}
\caption{(Color online) Trajectories of two S-matrix poles of a 1D uniform cavity when gain (Im$[n(x)]$) is added in different ways. In all cases squares indicate the passive cavity resonances $k^{(0)}_m$ and the diamonds mark the threshold lasing frequencies. (a,b) Gain is added uniformly to the whole cavity and to the left half of the cavity, respectively. In the former case one pole creates one lasing state, in the latter an anti-crossing in the complex plane of two poles generates an additional surface mode. (c-f) Gain ($n_i$) and loss ($-\tau n_i$) are added to the left and right halves respectively, at a ratio of $\tau=0.274,0.56,0.571,1$ ($\cal PT$-symmetric). For (c) $\tau= 0.274$ the poles hit the EP exactly, after which their vertical motions are reversed compared to the $\tau = 0$ case; the resulting surface mode frequency is shifted lower.  For (d) $\tau = 0.56$ the anti-crossing is almost at the real axis, and the surface and conventional modes (now no longer very different) have similar frequencies.  For (e) $\tau=0.571$ a inverse bifurcation occurs, the conventional and surface modes ``merge" and annihilate; for larger values or $\tau$, half the lasing modes of the uniformly pumped cavity are absent, and at the $\cal PT$-symmetric point ($\tau = 1$), there is a single CPA-laser mode for each pair of passive cavity modes, exactly half-way in between.
Other neighboring pairs behave similarly. }
\label{fig:antiX}
\end{center}
\end{figure}

Starting with the lossless cavity, when gain is added
(either uniformly or non-uniformly), the poles move upwards towards the real axis as we increase the gain (i.e. increase $n_i$) until the poles pass through the real axis at (in general) different values of $n_i$, corresponding to different thresholds for lasing (neglecting non-linear effects \cite{SPASALT}). For a high-Q cavity with uniformly distributed gain, the trajectories of the poles are almost vertical. Thus the lasing frequency can be well described by the real part of $k_m^{(0)}$, the (real) passive cavity frequencies, and the spatial distribution of the lasing mode will be similar to the passive cavity mode (inside the cavity). Moreover, in this case the poles continue to move upwards as the gain is increased beyond threshold and that pole will never cross the real axis again (see Fig.~\ref{fig:antiX}(a)) \cite{bibnote:poles}. Thus the lasing modes will be in one-to-one correspondence with the passive cavity modes.

The situation is very different for the case of a spatially inhomogeneous gain profile, however.  In this case the poles do not
move simply vertically, but in fact undergo an anti-crossing in the upper-half plane, which results in one of the two ``interacting" poles moving back down to the real axis and giving rise to a {\it second} lasing mode.  Thus, for this case one pole generates two different lasing modes at different values of the gain; however the second mode generated looks very different from the first one, and is one of the ``extra" surface modes (see Fig.~\ref{fig:antiX}(b)).

Another case of recent interest in which unconventional lasing modes appear is that of $\cal PT$-symmetric cavities.  These are optical cavities in which gain and loss are equally applied in such a manner that $n(-x) = n^* (x)$ \cite{PTrefs}.  The first $\cal PT$-symmetric optical systems were studied not in cavities, but in parallel balanced gain-loss waveguides \cite{PTrefs}.  More recently several authors have looked at cavities or heterojunctions \cite{mostafazadeh,Longhi,CPALaser}, and discovered a novel type of lasing transition, in which not only does a pole of the S-matrix reach the real axis, but a {\it zero} does so as well.
A zero of the S-matrix on the real axis corresponds to a ``coherent perfect absorber" (CPA): a cavity with loss tuned to absorb perfectly the time-reverse of the lasing mode when equivalent gain is added \cite{cpa_prl,cpa_science}.  For the $\cal PT$-symmetric laser (CPA-laser) one has the unique situation in which the cavity is strongly amplifying for one eigenmode of the S-matrix and strongly attenuating for the other.  In the CPA-laser, like the partial gain laser considered above, the lasing modes are not in one-to-one correspondence with the passive cavity resonances, but in the case of the CPA-laser {\it half} of the passive cavity resonances give rise to lasing modes and the other half do not, so there are {\it fewer} lasing modes, instead of extra modes. However it is obvious that one route to a CPA-laser is to start with a partial gain laser such as that of Fig. 1, with $L_G = L/2$, and simply add equal loss to the gain-free region.  Thus there should be some continuous change in the pole behavior to interpolate between the case of extra surface modes and the case of CPA-laser modes with half the density of the conventional lasing modes.

\begin{figure}
\begin{center}
\includegraphics[width=0.8\linewidth]{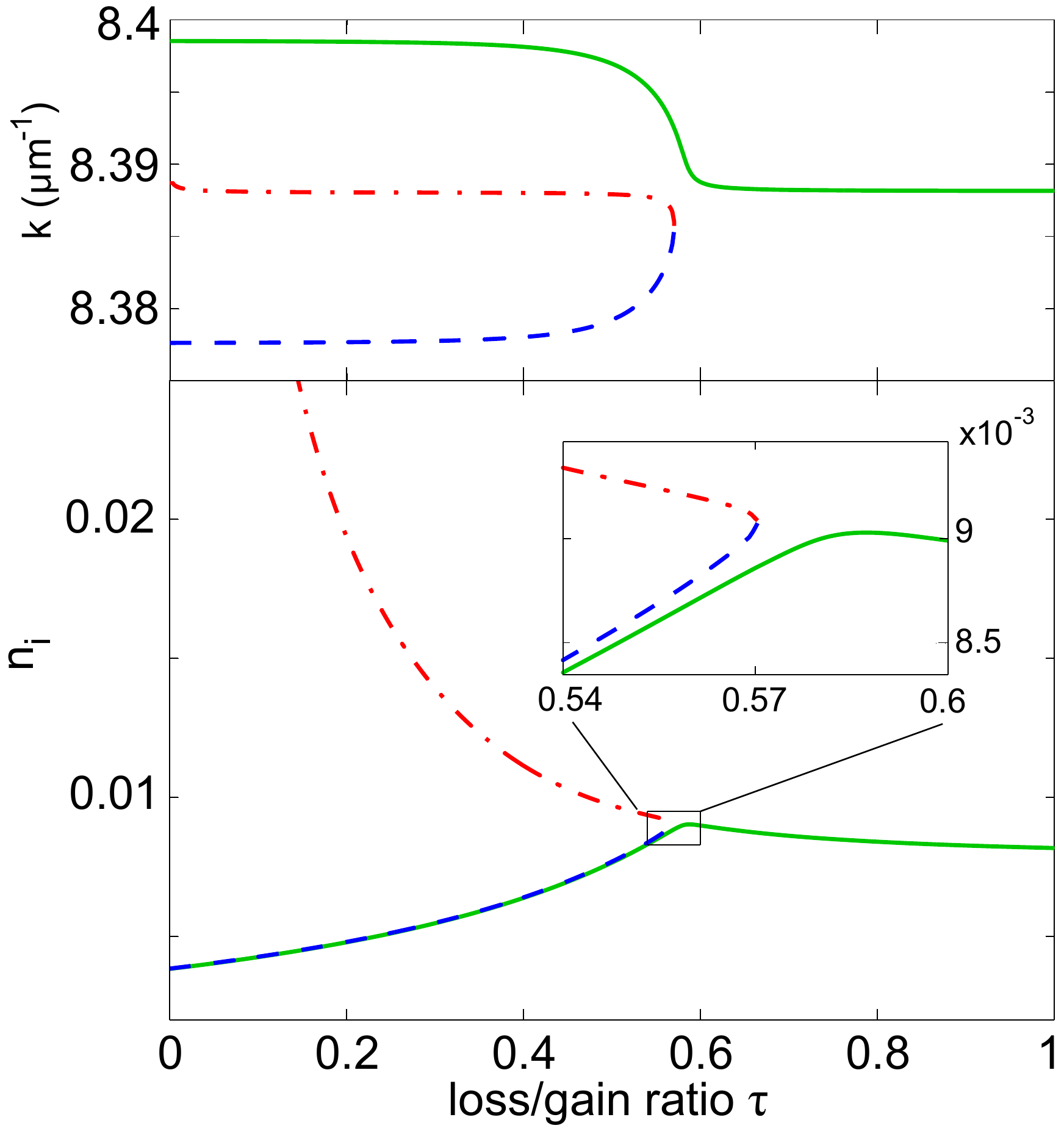}
\caption{(Color online) Evolution of threshold lasing frequency and gain as the loss/gain ratio $\tau$ increases from 0 (partial gain) to 1 ($\cal PT$-symmetric). The solid and dashed lines represent two passive cavity modes, and the dashed-dotted line indicates a surface mode.}
\label{fig:reduceLG}
\end{center}
\end{figure}

To find the connection between these CPA-laser states and the surface lasing modes, we gradually increase the loss/gain ratio $\tau$ from 0 (partial gain) to 1 ($\cal PT$-symmetric) and monitor the motion of the poles as the gain is increased at fixed $\tau$ (see Fig.~\ref{fig:antiX}(b-f)). We find that the anti-crossing in the upper-half plane of the two cavity resonances, which gives rise to the surface mode, occurs nearer to the real-axis as $\tau$ increases. Hence the pole which returns to the real axis and generates two modes at different values of the gain, now does so at closer and closer values of the frequency and gain, until, at a particular value of
$\tau \approx 0.571$, the pole motion reaches a point of tangency to the real axis and only a single mode is generated (see Fig.~\ref{fig:antiX}(d)).  This is another inverse bifurcation of the solution set; for infinitesimally larger $\tau$ this pole never crosses the real axis and both the surface and one of the conventional modes disappear; at $\tau =1 $ we reach the CPA-laser modes of the $\cal PT$-symmetric cavity, with half the modal density of the conventional lasing modes.  The behavior of the lasing
frequencies and thresholds through this transition are shown in Fig.~\ref{fig:reduceLG}.  We see that the
behavior of the surviving cavity mode after the bifurcation lies on a continuous curve with the surface mode.

\begin{figure}
\begin{center}
\includegraphics[width=0.8\linewidth]{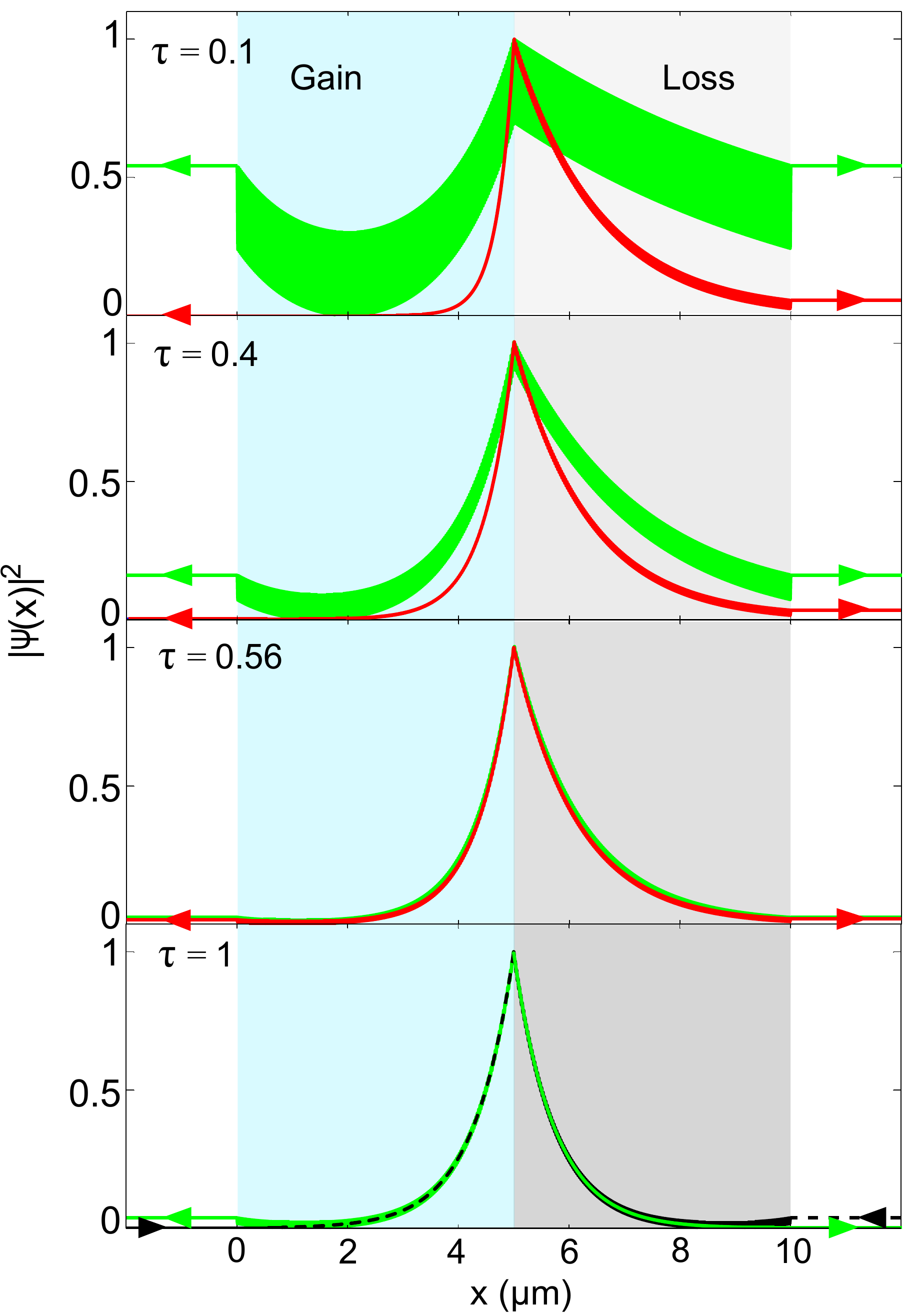}
\caption{(Color online) Mode profiles of the surface lasing mode and the surviving conventional lasing mode in Fig.~\ref{fig:reduceLG} with matching color schemes. Four values of $\tau$ are chosen: $0.2, 0.4, 0.56$ (just before the inverse bifurcation) and 1 ($\cal PT$-symmetric). The corresponding CPA mode (with incoming waves outside the cavity) in the $\cal PT$-symmetric case is given by the dashed black curve, which differs only slightly. The maximal intensities of all modes are normalized to unity. Individual oscillations cannot be resolved at the scale shown due to the large values of $n_1kL$.}
\label{fig:WF_largek}
\end{center}
\end{figure}

This variation of the lasing spectrum with $\tau$ is also reflected in the spatial profile of the modal intensities  (see Fig.~\ref{fig:WF_largek}).  For $\tau$ increasing from zero toward the bifurcation point, the paired conventional and surface modes begin to resemble each other more, until they become identical at the bifurcation point. Above the bifurcation, as already noted, the surviving conventional mode takes on the character of the surface mode which has disappeared.
Eventually, as $\tau \to 1$, the point of $\cal PT$-symmetry, this surviving mode becomes localized at the gain-loss boundary and corresponds to a CPA-laser mode described in \cite{CPALaser}.  Note that at the
$\cal PT$-symmetry point a {\it zero} of the S-matrix also sits on the real axis (not shown); this implies that there are modes of excitation of the cavity which will be perfectly absorbed, despite the fact that they sit at the lasing threshold \cite{CPALaser}. The spatial profiles of these CPA modes are obtained simply by reflecting the lasing modes around the gain-loss boundary: $|\Psi_{\text{CPA}}(x)| = |\Psi_{\text{Laser}}(-x)|$ (see Fig.~\ref{fig:WF_largek}). The  difference between the CPA and laser modes is subtle near the boundary of the gain and gain-free region, but becomes more substantial farther away.

The center of the avoided crossing regime in the complex plane is a branch point of two neighboring Riemann sheets, or an exceptional point (EP) \cite{EP}, where the S-matrix is defective (only has one eigenvector). The pole motion with varying $n_i$ accelerates as they approach the EP and decelerates once they are ``deflected" by the EP. Note that the vertical motions of the two poles after the deflection are different in Fig.~\ref{fig:antiX}(c-f) and in Fig.~\ref{fig:antiX}(b); the pole starting with the lower frequency moves downwards in the first two cases but moves upwards in the latter. This behavior suggests that the two poles must hit the EP simultaneously at an intermediate value of $\tau$. To find this value and identify the EP, we resort to the criterion $dk_m/dn_i \rightarrow \infty$. Using Eq.~(\ref{eq:M22=0}) we can write this criterion explicitly as
\bea
\tan[\theta_1 + n_1kL/2] &=& \pm i\sqrt{2}\left[1+\left(\frac{n_2}{n_1}\right)^2\right]^{-1/2},\\
\tan[\theta_2 + n_2kL/2] &=& \mp i\sqrt{2}\left[1+\left(\frac{n_1}{n_2}\right)^2\right]^{-1/2}.
\eea
By solving the real and imaginary parts of these two equations, we obtain not only the position of the EP $k = 8.388 + 0.012i\, \mu{m}^{-1}$ but also the required threshold gain $n_i=0.0113$ and loss/gain ratio $\tau=0.274$ for the poles to reach the EP (Fig.~\ref{fig:antiX}(c)). Note that the EP does not cause the annihilation behavior discussed above directly. However, its physical presence can be observed by an alternative method \cite{TU}.

\section{Conclusion}

We have discussed a class of highly localized lasing modes created by a spatially inhomogeneous gain profile. Both their frequencies and spatial profiles are very different from the passive cavity modes, which are the origin of the lasing modes within conventional laser theory. Such surface modes occur more prominently in low-Q lasers, such as random lasers, but have nothing to do physically with randomness; they occur as well in the simple, uniform, low-Q dielectric cavities studied above.
We have identified the physical origin of these surface modes as the transmission resonances of the gain-free region. Using an S-matrix approach we have shown the connection between the surface modes and the lasing modes in $\cal PT$-symmetric cavities. Our study suggests the possibility to achieve anisotropic emission in lasers by tailoring the gain profile rather than changing the cavity shape or mirror reflectivities. Due to the highly asymmetric spatial profile of the surface modes, they may find applications in optical switching and sensing \cite{SMapp}.

\section{Acknowledgements}

This work was partially supported by NSF Grant No. DMR-0908437, seed funding from the Yale NSF-MRSEC (DMR-0520495), Swiss NSF Grant No. PP00P2-123519/1 and the Vienna Science Fund (WWTF). We thank Jonathan Andreasen, Matthias Liertzer and Hui Cao for helpful discussions.


\begin{thebibliography}{100}
\bibitem{bibnote0} We will assume all regions containing the gain medium are pumped.
\bibitem{bibnote:poles} The poles eventually move down towards the origin when the gain becomes unphysically large, but they will not cross the real axis again.
\end{thebibliography}
\end{document}